\documentclass{aa} 
\usepackage{graphicx}
\usepackage{txfonts}
\usepackage{natbib}
\usepackage{capt-of}
\usepackage{textcomp}
\usepackage{epstopdf}
\usepackage[version=3]{mhchem}
\usepackage[usenames,dvipsnames]{xcolor}
\usepackage{gensymb}
\begin{document} 

\renewcommand{\thefootnote}{\alph{footnote}}
\renewcommand{\thefootnote}{\fnsymbol{footnote}}

\title{Methanol ice co-desorption as a mechanism to explain cold methanol in the gas-phase}

\author{N.~F.~W.~Ligterink
       \inst{1,2}, 
       C.~Walsh \inst{2,3},
       R.~G.~Bhuin \inst{1,4}, 
       S.~Vissapragada \inst{2,5}, 
       J.~Terwisscha van Scheltinga \inst{1,2}
       \and H.~Linnartz \inst{1}
       }

\institute{Raymond and Beverly Sackler Laboratory for Astrophysics, Leiden Observatory, 
          Leiden University, PO Box 9513, 2300 RA Leiden, The Netherlands\\      
          \email{ligterink@strw.leidenuniv.nl}
          \and
          Leiden Observatory, Leiden University, PO Box 9513, 2300 RA Leiden, The Netherlands   
          \and
          School of Physics and Astronomy, University of Leeds, Leeds, LS2 9JT, UK\\  
          \email{c.walsh1@leeds.ac.uk}
          \and
          Lehrstuhl für Physikalische Chemie II, Friedrich-Alexander-Universität Erlangen-                       Nürnberg, Egerlandstr. 3, 91058 Erlangen, Germany
          \and
          Columbia Astrophysics Laboratory, Columbia University, New York, NY 10027, USA}

\date{Received $\dots$ / Accepted $\dots$}

\titlerunning{Methanol codesorption with CO}
\authorrunning{Ligterink et al.}

 
\abstract
{Methanol is formed via surface reactions on icy dust grains. Methanol is also detected in the gas-phase at temperatures below its thermal desorption temperature and at levels higher than can be explained by pure gas-phase chemistry. The process that controls the transition from solid state to gas-phase methanol in cold environments is not understood.}
{The goal of this work is to investigate whether thermal CO desorption provides an indirect pathway for methanol to co-desorb at low temperatures.}
{Mixed \ce{CH3OH}:CO/\ce{CH4} ices were heated under UHV (ultra-high vacuum) conditions and ice contents are traced using RAIRS (reflection absorption IR spectroscopy), while desorbing species were detected mass spectrometrically. An updated gas-grain chemical network was used to test the impact of the results of these experiments. The physical model used is applicable for TW Hya, a protoplanetary disk in which cold gas-phase methanol has recently been detected.}
{Methanol release together with thermal CO desorption is found to be an ineffective process in the experiments, resulting in an upper limit of $\leq7.3 \times 10^{-7}$ \ce{CH3OH} molecules per CO molecule over all ice mixtures considered. Chemical modelling based on the upper limits shows that co-desorption rates as low as 10$^{-6}$ \ce{CH3OH} molecules per CO molecule are high enough to release substantial amounts of methanol to the gas-phase at and around the location of the CO thermal desorption front in a protoplanetary disk. The impact of thermal co-desorption of \ce{CH3OH} with CO as a grain-gas bridge mechanism is compared with that of UV induced photodesorption and chemisorption. }
{}


\maketitle


\section{Introduction}
\label{Introduction}

Methanol is one of the smallest complex organic molecules \citep[COMs,][]{herbstdishoeck2009} observed in the interstellar medium (ISM). It has been identified in both the gas-phase and solid-state towards many types of sources, among which are dark clouds, hot cores, protostellar envelopes, protoplanetary disks and comets \citep{ball1970,grim1991,bockeleemorvan1991,bisschop2007,bottinelli2010,taquet2015,boogert2015,walsh2016}. Gas-phase reactions account only for a small fraction of methanol production \citep{geppert2005,garrod2006b}. It has been demonstrated that methanol ice efficiently forms by sequential hydrogenation of CO on interstellar dust grain analogues \citep[][and references therein]{watanabe2002,fuchs2009,linnartz2015}. Evidence for this formation pathway in the ISM is supported by infrared (IR) observations of interstellar ices, which show that CO and methanol are both abundant ice species that are intimately mixed \citep{pontoppidan2003,cuppen2011,penteado2015}.

The release of methanol from ice mantles on dust grains to the gas-phase is generally thought to occur either via thermal desorption or photodesorption, depending on the physical conditions governing the local environment. For example, young stellar objects (YSOs) enrich the gas-phase with warm methanol released by thermal desorption in their hot core or corino phase \citep[e.g.][]{turner1991,vandishoeck1995,bisschop2007,taquet2015}. Photodesorption of methanol ice has been invoked to explain gas-phase abundances of rotationally cold methanol in cold ($\ll 100$~K) environments such as dark clouds \citep{friberg1988}.  
\\
\indent
The photodesorption mechanism, whereby a VUV (vacuum-ultraviolet) photon impinges on solid-state molecules and induces a release to the gas-phase, is well established for CO ice. Quantification of this process has been realised mainly in the laboratory \citep{oberg2007,munozcaro2010,fayolle2011,chen2014,paardekooper2016} and is also theoretically supported \citep{hemert2015}. It has been used to interpret recent observations of a disk showing a double snowline \citep{oberg2015b}. Experimental and theoretical evidence of intact water photodesorption is also available \citep{westley1995,andersson2005,anderssondishoeck2008,oberg2009c,arasa2015}. Methanol photodesorption has not been as rigorously studied. \citet{oberg2009a} indirectly inferred a photodesorption rate for pure methanol ice of $2.1 \times 10^{-3}$ molecules photon$^{-1}$ from fitting multiple components to the rate of methanol ice loss. \citet{cruzdiaz2016} recently revisited the broad-band VUV irradiation of pure methanol ice including direct detection of the photodesorption products via mass spectrometry. Only products of VUV ice processing were directly detected (\ce{H2}, CO, and \ce{CH4}) leading to an upper limit of $< 3 \times 10^{-5}$ molecules photon$^{-1}$ photodesorption of intact methanol. At the same time \citet{bertin2016} performed wavelength dependent experiments on methanol and methanol:CO mixed ice. In both the pure and mixed ice the role of dissociative photodesorption was studied and the photodesorption rate of intact methanol was found to be low \citep[in agreement with][]{cruzdiaz2016}. For a pure methanol ice this is in the order of 1-2$\times$10$^{-5}$ molecules photon$^{-1}$, whereas for the mixed ice an upper limit of $\leq$3$\times$10$^{-6}$ molecules photon$^{-1}$ was found. The resulting methanol photodesorption rates are several orders of magnitude lower than the typical yields of $10^{-2}$-$10^{-3}$ molecules photon$^{-1}$ for CO and \ce{H2O} photodesorption and are too low to explain gas-phase abundances of methanol in cold regions of the ISM. Also, the result of the mixed ice is interesting because it shows that indirect photodesorption, known to be efficient for the N$_2$:CO case \citep{bertin2013}, is an inefficient process in the case of methanol:CO mixtures. Therefore to explain the presence of rotationally cold methanol other transfer mechanisms from the solid-state to the gas-phase need to be investigated.

One such option is chemical or reactive desorption. In this mechanism two fragments, usually radicals or ions, react with each other to form the product species and the excess energy from the reaction is available to release the product from the solid-state to the gas-phase. This mechanism was investigated by \citet{martindomenech2016} for methanol. In VUV irradiated \ce{H2O}:\ce{CH4} ices they concluded on the chemi-desorption of photo-produced formaldehyde at $\approx 4.4 \times 10^{-5}$ molecules photon$^{-1}$;  chemi-desorption of methanol ice was not found, even though the experiments showed \ce{CH3OH} formation.

Another possible transfer mechanism is thermal co-desorption \citep{sandford1988,oberg2005,fuchs2006,brown2007,martindomenech2014,burke2015,burkebrown2015,urso2017}. This term is applied to any thermal release of a molecule to the gas-phase induced by a second matrix ice species. Most cases involve species that are trapped above their respective desorption temperature in a matrix with a higher desorption temperature than said species. These species release partially to the gas-phase at their regular desorption temperature and co-desorp when the matrix undergoes a phase change (for example the change of amorphous solid water to crystalline water) or when the matrix itself begins to thermally desorb \citep[e.g.][]{bar-nun1985}. In both cases, co-desorption is driven by the fact that the molecule is initially hindered to thermally desorb because of its matrix environment. The opposite of the previous case is co-desorption of a molecule below its desorption temperature, that is, when the matrix has a lower desorption temperature than the molecule, and the matrix carries the less volatile species with it when desorbing. This specific kind of co-desorption is the subject of the present study and further mention of the term co-desorption refers to desorption of a molecule below its desorption temperature induced by the thermal release of matrix species. 

It is well established that methanol ice forms in situ on ice mantles on cold dust grains via the sequentional hydrogenation of CO ice \citep{watanabe2002,fuchs2009}. We therefore investigate the potential release of methanol to the gas-phase with a thermally desorbing CO matrix at low temperatures. As noted above, recent photodesorption and reactive desorption experiments suggest that methanol does not come off intact in either case \citep{bertin2016,cruzdiaz2016,martindomenech2016}. The aim here is to test if the proposed co-desorption mechanism can explain the presence of rotationally cold gas-phase methanol in the ISM. Laboratory experiments are discussed that study the thermal desorption dynamics of CO:\ce{CH3OH} ice mixtures and the results are tested in astrochemical models. The models adopted are representative of the physical conditions in the protoplanetary disk around TW~Hya in which cold gas-phase methanol was recently detected with ALMA \citep{walsh2016}. This was the first detection of methanol in a protoplanetary disk and is a particularly interesting test case for thermal co-desorption given the range of temperatures in this or similar objects.

This paper is organised in the following way. The experimental set-up and procedure are described in Sect. \ref{sec.exp}, that also summarises the experimental results. In Sect. \ref{sec.models}  the astrochemical model is described and used to interpret the experimental findings. The conclusions are presented in Sect. \ref{conclusion}.


\section{Experimental setup and results}
\label{sec.exp}

\subsection{Laboratory set-up and protocols} 

\subsubsection{CryoPAD2}
\label{sec.cp2}

All laboratory measurements in this paper were carried out on the Cryogenic Photoproduct Analysis Device 2 (CryoPAD2), a recently upgraded setup to study VUV induced processes in interstellar ice analogues. In short, this setup consists of a main chamber at oil-free, ultra high vacuum (UHV, <$1\times 10^{-10}$~mbar). At its centre a cryogenically cooled gold coated reflective surface was mounted, which can be cooled to a lowest temperature of 12~K. Gasses were prepared in a gas mixing line and deposited on the cryogenically cooled surface to form an ice layer, using  a leak valve connected to a nozzle positioned in front of deposition zone. A Lakeshore Model 350 temperature controller controlled the feedback loop between a thermocouple and heating wire to set the base temperature and temperature ramp on the substrate with a relative accuracy better than $\pm 1$~K, and absolute accuracy not exceeding $\pm 2$~K. The composition of the deposited ices were analysed in situ by Reflection Absorption InfraRed Spectroscopy (RAIRS), by impinging an IR beam on the substrate at an angle of 83$\degree$ with respect to the normal. A resolution of 2 cm$^{-1}$ was used in all measurements. Gas-phase species were analysed with a Hiden 3F RGA quadrupole mass spectrometer (QMS), which was connected to the temperature controller and recorded the temperature at the same time. This QMS faced the gold surface directly. Mass fragmentation patterns of desorbing species can be linked to characteristic temperatures in order to record temperature programmed desorption (TPD) mass signals.
 
\subsubsection{Experimental protocol}
\label{protocol}

\begin{table}
\centering
\caption[]{IR positions and transmission band strengths of CO, methanol, methane and isotopologues}
\label{tab.band} 
\begin{tabular}{c c c c}
\hline
\noalign{\smallskip}
Molecule & Mode & Position & Transmission band strength \\
& & cm$^{-1}$ & $\times10^{-17}$ cm molecule$^{-1}$ \\
\noalign{\smallskip}
\hline
\noalign{\smallskip}
CO & CO str. & 2138 & 1.12$^{a}$\\
$^{13}$CO & CO str. & 2092 & 1.32$^{a}$ \\
$^{13}$\ce{CH3OH} & CO str. & 1020 & 1.07$^{a}$ \\
\ce{CH4} & deform. & 1302 & 0.97$^{b}$ \\
\noalign{\smallskip}
\hline
\noalign{\smallskip}
\end{tabular}  
\\
\emph{\rm Notes. $^{a}$\citet{bouilloud2015}, $^{b}$\citet{gerakines2015}}
\end{table}

Mixtures of carbon monoxide (Linde Gas, 99.997$\%$) or methane (Linde Gas, 99.995$\%$) and $^{13}$C-methanol (Sigma-Aldrich, 99\%) were prepared in the mixing line. $^{13}$\ce{CH3OH} has a unique mass at $m/z$ 33, which does not overlap with the masses of the $^{12/13}$C and/or $^{16/18}$O carbon monoxide isotopologues, nor with potential contaminants like molecular oxygen ($^{16}$O$_{2}$ at $m/z$ 32, $^{16}$O$^{18}$O at $m/z$ 34). Methanol was purified in a number of freeze-pump-thaw cycles before use. The mixing ratio was determined with a gas-independent gauge. 

Deposition on the substrate occurs at 15~K, after which an IR spectrum was taken to verify and characterise the composition of the ice. The system is given time to pump any residual gasses of the deposition until a pressure of $1 \times 10^{-10}$ mbar or better is reached, in order to have good baseline conditions. Next, a temperature ramp, typically 10 K min$^{-1}$, was set to heat the deposited sample. The high heating ramp was chosen in order to give the ice less time to undergo structural changes and make it more likely that methanol will co-desorb with CO. IR spectra are continuously recorded in order to trace ice changes due to thermal processing. In parallel, the QMS records the (co-)desorption of molecules from the sample, focusing mainly on $m/z$ 16 and 33.

Recorded mass spectra were corrected for the QMS response function. Since the QMS detector can saturate at high signal intensities, a good alternative approach is to trace desorption of a certain molecule at a minor fragmentation channel instead of its main channel. Particularly for CO, of which large amounts desorb in the experiments, it is better to trace this molecule at $m/z$~=~16, because the O$^{+}$ fragment signal is $\sim$60 times less intense in the signal than that of the main CO$^{+}$ fragment channel at $m/z$~=~28.

Column densities ($N_{\rm species}$) were determined from the IR spectra by the equation:

\begin{equation}
\label{eq.coldens}
N_{\rm species} = \frac{1.1}{3.4}{\rm ln(10)}\frac{\int_{\rm band}log_{\rm 10}\left(\frac{I_{\rm 0}(\tilde{\nu})}{I(\tilde{\nu})}\right) d\tilde{\nu}}{A'_{\rm }},
\end{equation}

where $\int_{\rm band}log_{\rm 10}\left(\frac{I_{\rm 0}(\tilde{\nu})}{I(\tilde{\nu})}\right) d\tilde{\nu}$ is the absorbance band area, with $I_{\rm 0}(\tilde{\nu})$ and $I(\tilde{\nu})$ respectively being the flux received and transmitted by the sample. $A'_{\rm band}$ is the apparent band strength and $\frac{1.1}{3.4}$ a RAIRS scaling factor explained below. Band strengths can be different in a RAIRS set-up with respect to transmission spectroscopy, due to the difference in path length. Therefore, additional band strength determinations have been performed. In our experiments, for CO the equivalent of one monolayer (ML) was determined mass spectrometrically from TPD experiments and correlated to the CO stretch absorption signal in the IR. By assuming a column density of 10$^{15}$ molecules cm$^{-2}$, the set-up specific RAIRS band strength for CO was found to be 3.4$^{+0.5}_{-0.5} \times 10^{-17}$ cm molecule$^{-1}$. Under the assumption that other set-up specific band strengths scale in the same way as CO, a factor of $\frac{1.1}{3.4}$ is used to adapt transmission band strengths taken from literature \citep{bouilloud2015}. The bands used for analysis and their apparent band strengths in transmission are listed in Table \ref{tab.band}. The total $^{12+13}$CO column density is determined by multiplying $N$($^{13}$CO) by (90+1), where 90 is the $^{12}$CO/$^{13}$CO isotope ratio. The $^{12}$CO band is deemed less reliable for column density determination because high intensity effects make a baseline fit more difficult and because it is susceptible to non-linear RAIRS effects \citep{teolis2007}. 

\subsubsection{Methanol co-desorption rate determination}
\label{sec.co-des_rate}

\begin{table}
\centering
\caption[]{Fragmentation patterns of CO and $^{13}$\ce{CH3OH} upon 70~eV electron impact ionisation.}
\label{tab.frag} 
\begin{tabular}{c c c c}
\hline
\noalign{\smallskip}
Mass & CO & $^{13}$\ce{CH3OH} & \ce{CH4} \\
(amu) & & \\
\noalign{\smallskip}
\hline
\noalign{\smallskip}
12 & 0.04268 & - & 0.01686 \\
13 & - & 0.00081 & \textcolor{red}{0.0474} \\
14 & - & 0.00245 & 0.09062 \\
15 & - & 0.00690 & 0.39404 \\
16 & \textcolor{red}{0.01580} & 0.05059 & 0.44375 \\
28 & 0.92936 & - & - \\
29 & 0.01115 & 0.01874 & -\\
30 & - & 0.18205 & -\\
31 & - & 0.02650 & -\\
32 & - & 0.40824 & -\\
33 & - & \textcolor{red}{0.30372} & -\\
\noalign{\smallskip}
\hline
\noalign{\smallskip}
\end{tabular}  
\\
\emph{\rm Notes. Fragmentation pattern and intensities based on NIST data. The most commonly used channels are highlighted in red.}
\end{table}

The co-desorption rate of methanol released in tandem with the thermal release of CO (or \ce{CH4}), $R_\mathrm{methanol}$, is calculated from the TPD spectrum using the following relation, 

\begin{equation}
\label{eq.yield}
R_\mathrm{methanol} = \frac{\phi_\mathrm{m/z,CO}}{\phi_\mathrm{m/z,methanol}} \frac{\sigma_\mathrm{CO}}{\sigma_\mathrm{methanol}}  \frac{A_\mathrm{methanol}}{A_\mathrm{CO}},
\end{equation}

where $\phi$ is the fragmentation fraction of CO and methanol at a specific mass, $\sigma$ the total electron impact ionisation cross section of CO (or methane) and methanol at 70~eV and $A$ the integrated QMS signal of CO or methanol. $\sigma_{\mathrm{CO}}$ is given as 2.44$\AA^{2}$ \citep{hudson2003b} and $\sigma_\mathrm{methanol}$ is 4.44$\AA^{2}$ \citep{hudson2003a}. For methane it is 3.93$\AA^{2}$ \citep{nishimuratawara1994}. The mass fragmentation patterns of CO, \ce{CH4}, and $^{13}$\ce{CH3OH} are given in Table \ref{tab.frag}. 

\subsection{Experimental results}
\label{sec.exp_res}

\begin{table*}
\centering
\caption[]{Upper limit co-desorption rates (R) for a systematic set of different experiments.}
\begin{tabular}{c c c c c c c c}
\hline
\noalign{\smallskip}
Methanol & CO$^{a}$ & CH$_{4}$ & Ratio & Type   & Heating ramp   & $T_{\rm peak}$(CO) & $R_\mathrm{methanol}$ \\
(ML) & (ML) & (ML) &       &            & (K min$^{-1}$) & (K) & (MeOH CO$^{-1}$) \\
\noalign{\smallskip}
\hline
\noalign{\smallskip}
5.7 & 137.4 & - & $\sim$1:24 & mixed & 1 & 32.3 & $\leq3.3 \times 10^{-6}$ \\
2.9 & 52.2 & - & $\sim$1:18 & mixed & 10 & 33.5 & $\leq2.6 \times 10^{-6}$ \\
8.7 & 119.7 & - & $\sim$1:14 & mixed & 10 & 35.1 & $\leq1.1 \times 10^{-6}$ \\
11.4 & 123.3 & - & $\sim$1:11 & mixed & 10 & 34.9 & $\leq1.9 \times 10^{-6}$ \\
24.6 & 126.9 & - & $\sim$1:5 & mixed & 10 & 35.4 & $\leq7.3 \times 10^{-7}$ \\
31.2 & 30.5 & - & $\sim$1:1 & layered & 10 & 32.5 & $\leq3.2 \times 10^{-5}$ \\
13.1 & 21.0 & - & $\sim$3:5 & mixed on CO & 10 & 32.8 & $\leq6.8 \times 10^{-6}$ \\
\noalign{\smallskip}
\hline
\noalign{\smallskip}
3.9 & - & 31.9 & $\sim$1:8 & mixed & 10 & 41.2$^{b}$ & $\leq3.2 \times 10^{-5,b}$ \\
\noalign{\smallskip}
\hline
\noalign{\smallskip}
\end{tabular} 
\label{tab.explist} 
\\
\emph{\rm Notes. $^{a}$Determined by the $^{13}$CO band multiplied by 91 to retrieve the total $^{12+13}$C column density. $^{b}$In the experiment making use of methane, $T_{\rm peak}$ is given for CH$_{4}$ and $R_\mathrm{methanol}$ is units of MeOH CH$_{4}^{-1}$.}
\end{table*}

Most co-desorption experiments conducted in this work have been performed with $^{13}$\ce{CH3OH}:CO mixed ices, but layered ices ($^{13}$\ce{CH3OH} deposited on top of a layer of CO) have been investigated as well. A variety of thicknesses and mixing ratios have been used. A full list of the performed experiments is given in Table \ref{tab.explist}.

\begin{figure}
\centering
\includegraphics[width=\hsize]{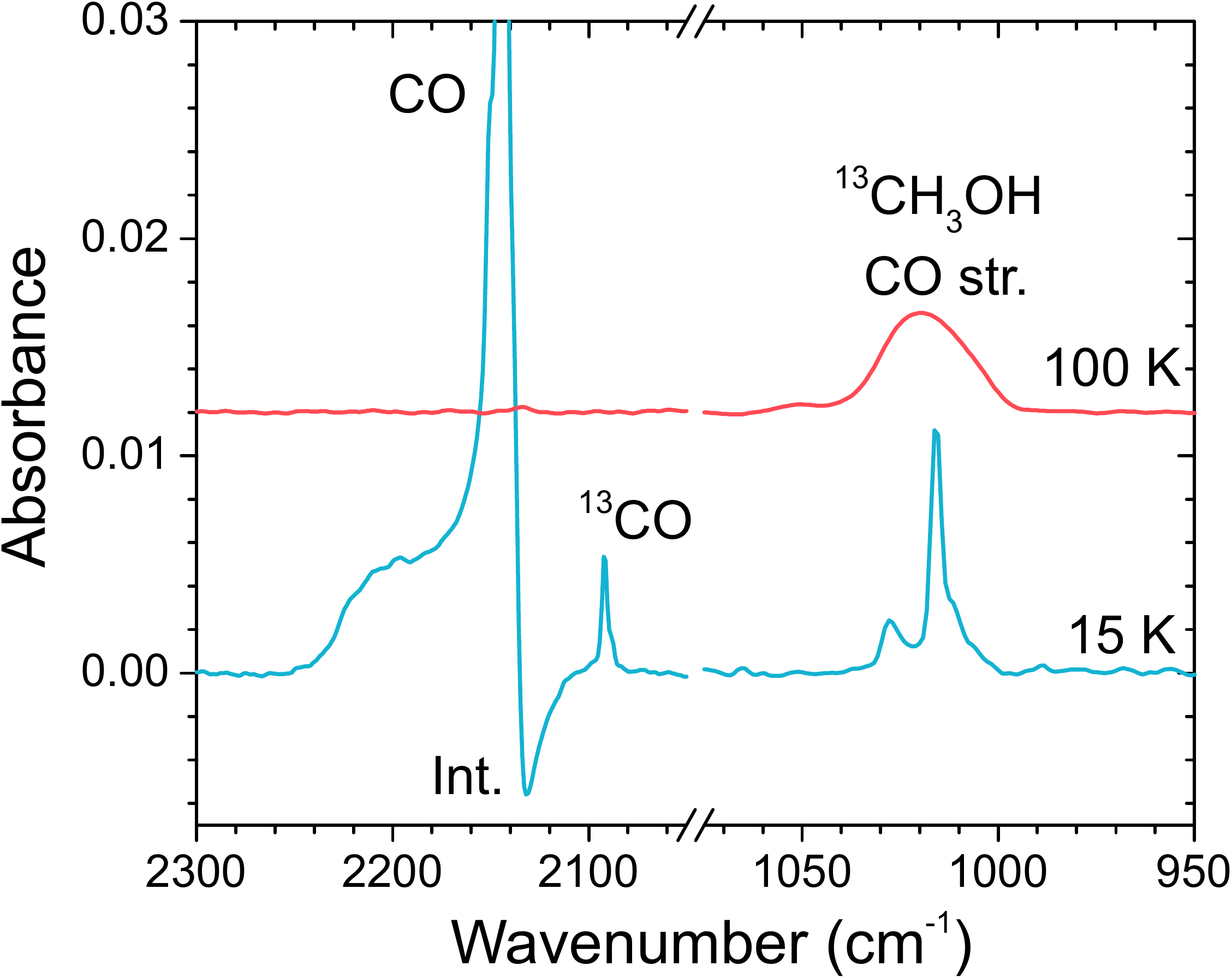}
\caption{Representative IR spectrum taken after deposition at 15~K (blue) of the 1:24 $^{13}$\ce{CH3OH}:CO mixture and during the TPD at 100~K (red), showing the main IR features of CO and $^{13}$\ce{CH3OH}. For the methanol CO stretch mode, the $^{13}$CO and CO peak (cut-off due to its high peak intensity) are visible. Between the two CO peaks an artefact, labelled int., is visible caused by the high intensity of the CO peak.}
\label{fig.IR_overall}
\end{figure}

A typical IR spectrum taken after deposition is shown in Fig. \ref{fig.IR_overall} for the 1:24 mixed $^{13}$\ce{CH3OH}:CO experiment. At 1015 cm$^{-1}$ the CO stretch mode of methanol is visible. We note that this mode has two components. The strong component is associated with monomeric methanol, methanol that is largely isolated in the CO matrix. The smaller, blue-shifted component is mainly caused by methanol clusters, but minor contributions of $^{12}$CH$_{3}$OH impurities in the sample cannot be excluded. Depending on the mixing ratio, ice thickness and temperature, the profile of this band can change from fully monomeric to clustered methanol. At 2092 cm$^{-1}$ the $^{13}$CO peak is visible and next to it the intense CO stretching mode at 2143 cm$^{-1}$. Between the two CO isotope peaks an artefact is visible that is caused by the high intensity of the CO peak. The same figure also shows the IR spectrum at 100~K during the TPD. CO has desorbed at this point, but methanol is still present. Its peak shape has changed, however, due to the removal of the CO matrix. Table \ref{tab.explist} lists the ice column densities of all experiments.

\begin{figure}
\centering
\includegraphics[width=\hsize]{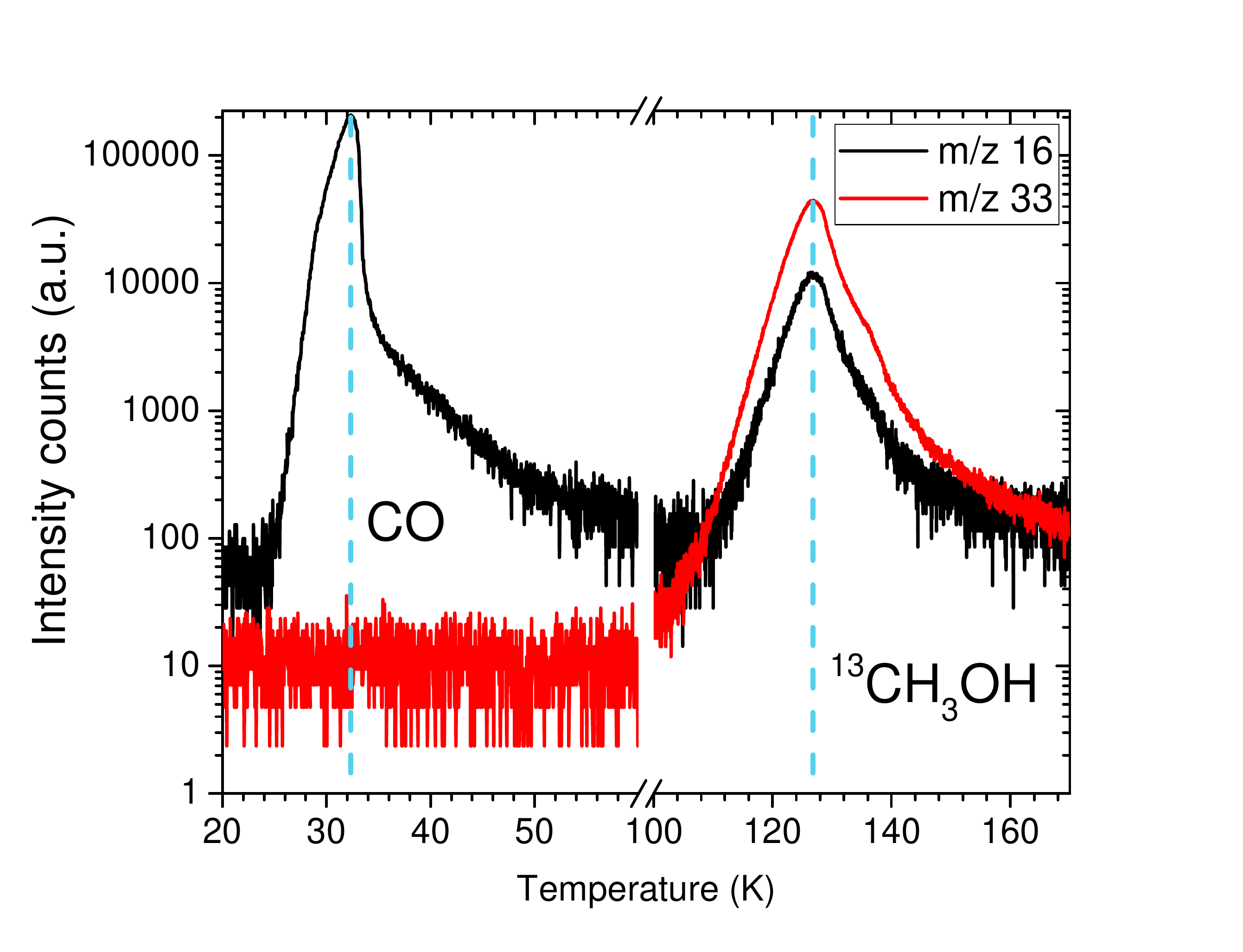}
\caption{Desorption profile of the 1:24 $^{13}$\ce{CH3OH}:CO mixture, heated at 1 K min$^{-1}$. CO is traced by $m/z$ 16 (black) and desorbs just above 30~K, while $^{13}$C-methanol is traced by $m/z$ 33 (red) and desorbs around 130~K. The presence of $m/z$ 16 at the methanol desorption peak is due to the $^{13}$\ce{CH3} (= 16 amu) fragment.}
\label{fig.des_overall}
\end{figure}

From the same experiment, a typical desorption pattern is shown in Fig. \ref{fig.des_overall}. CO is traced at $m/z$ 16, while $^{13}$\ce{CH3OH} is traced by $m/z$ 33. CO desorbs around 30~K \citep[e.g.][]{acharyya2007}, while bulk methanol desorption occurs around 130~K. The $m/z$ 16 signal around 130~K is the $^{13}$CH$_{3}$ fragment of the methanol fragmentation pattern, but can partially also be caused by trace amounts of CO trapped in the methanol ice and releasing upon methanol desorption. No release of $m/z$ 33 is seen around 30~K.

\begin{figure}
\centering
\includegraphics[width=\hsize]{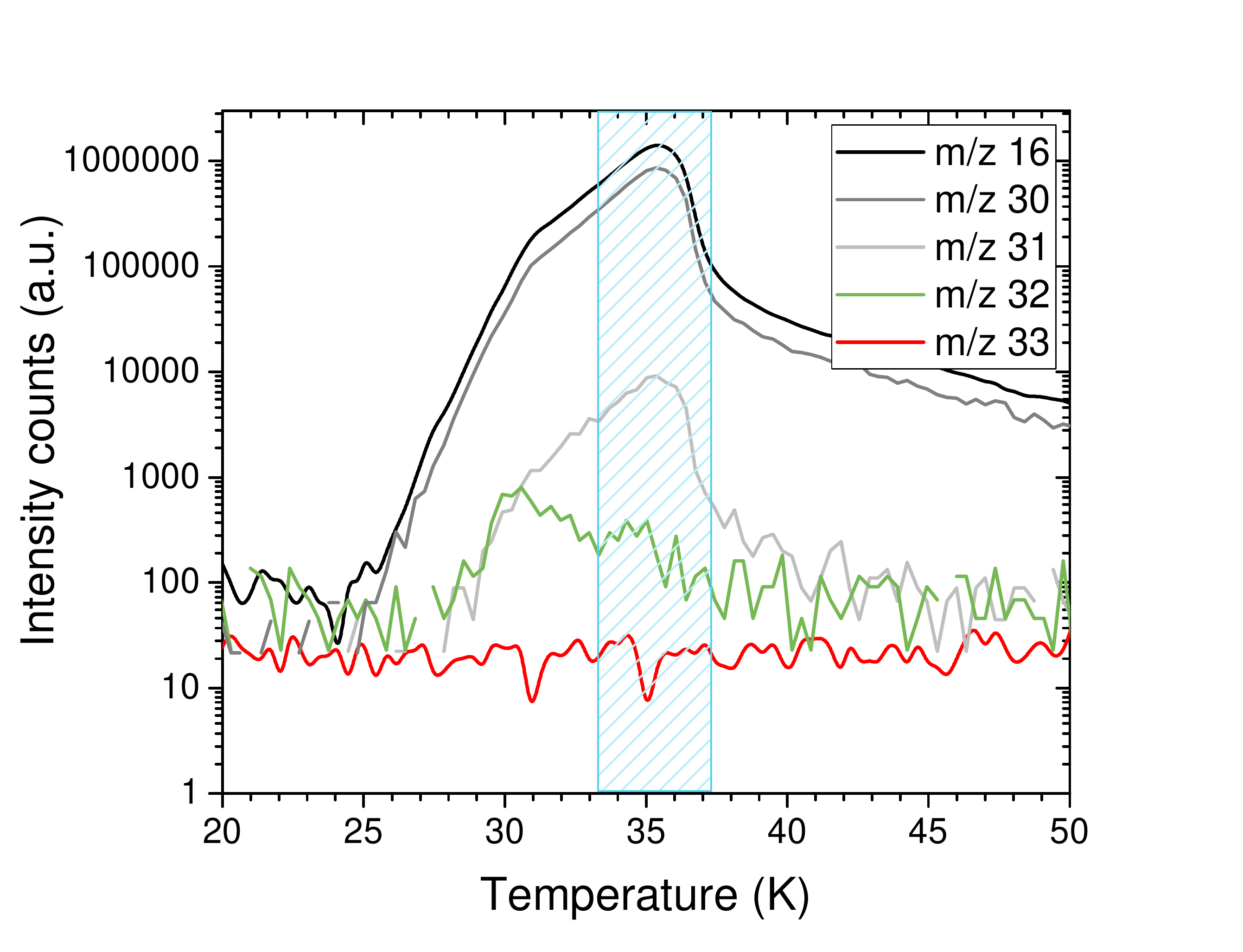}
\caption{Close-up of the CO desorption peak (shaded blue) for the 1:5 $^{13}$\ce{CH3OH}:CO mixed ice. $m/z$ 16 (black) traces CO, while $m/z$ 30 (dark grey) and 31 (light grey) trace C$^{18}$O and $^{13}$C$^{18}$O, respectively. The signal in $m/z$ 32 (green) is caused by minor traces of $^{16}$O$_{2}$. $m/z$ 33 (red) is associated with $^{13}$\ce{CH3OH}. We note the logarithmic scale.}
\label{fig.lowest}
\end{figure}

Figure \ref{fig.lowest} shows a close-up of the CO desorption peak, traced at $m/z$ 16, for the 1:5 $^{13}$\ce{CH3OH}:CO mixed ice experiment. The C$^{18}$O and $^{13}$C$^{18}$O isotopes are shown as well at $m/z$ 30 and 31, respectively. At the CO desorption peak no increase of $m/z$ 33 is seen. An increase in signal of $m/z$ 32 is seen, but at a ratio of $m/z$ (32/33) > 10, much larger than the same fragment ratio of $\sim$1.3 in the $^{13}$\ce{CH3OH} fragmentation pattern (see Table \ref{tab.frag}). Therefore, it is unlikely that $^{13}$\ce{CH3OH} is co-desorbing. $m/z$ 32 is likely tracing small quantities of O$_{2}$ contamination. Using the formalism described in Sect. \ref{sec.co-des_rate}, an upper limit co-desorption rate can be determined from the CO peak and $^{13}$\ce{CH3OH} $m/z$ 33 baseline; the inferred limit in this specific experiment is $R_\mathrm{methanol} \leq7.3 \times 10^{-7}$ \ce{CH3OH} CO$^{-1}$.

In none of the experiments, including the layered ones, did we find $m/z$ 33 desorbing simultaneously with CO. This leads to a series of upper limit co-desorption rates listed in Table \ref{tab.explist}. The above mentioned upper limit is the most constraining, but generally the upper limits are found to be in the order of a few $\times 10^{-6}$ \ce{CH3OH} CO$^{-1}$.

\begin{figure}
\centering
\includegraphics[width=\hsize]{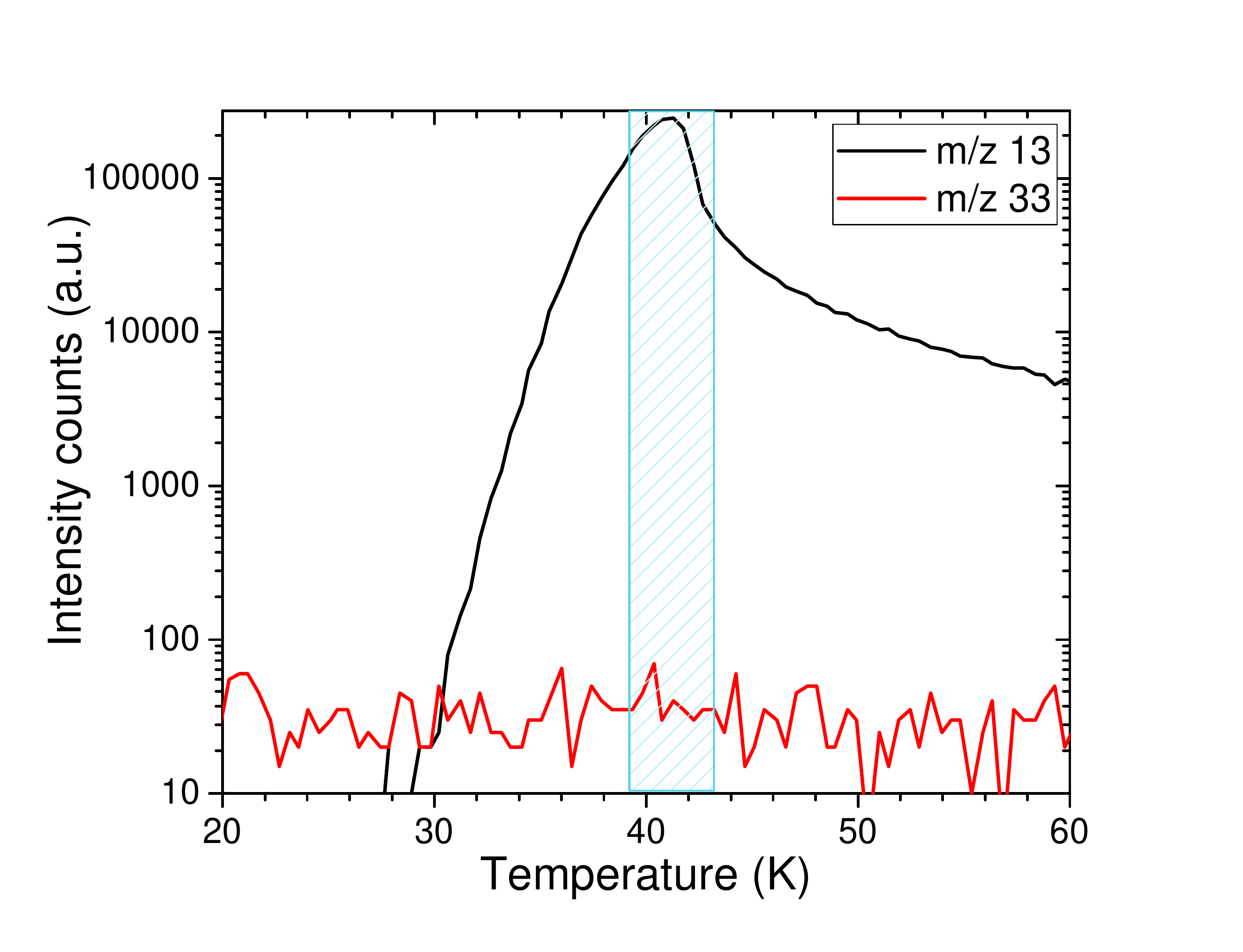}
\caption{Close-up of the \ce{CH4} desorption peak (shaded blue) for the 1:8 $^{13}$\ce{CH3OH}:\ce{CH4} mixed ice. $m/z$ 13 (black) traces methane, while $m/z$ 33 (red) is associated with $^{13}$\ce{CH3OH}. We note the logarithmic scale.}
\label{fig.ch4}
\end{figure}

To test whether other low volatility molecules can induce co-desorption of methanol, a TPD experiment was run with a $^{13}$\ce{CH3OH}:\ce{CH4} mixture. Methane has a comparable desorption temperature to CO of roughly 40~K \citep{collings2004}. Figure \ref{fig.ch4} shows a close-up of the methane desorption peak around 41 K, traced by $m/z$ 13. The methanol, traced by $m/z$ 33, is again not co-desorbing. In a similar way as described before, for this experiment an upper limit $R_\mathrm{methanol} \leq3.2 \times 10^{-5}$~\ce{CH3OH}~\ce{CH4}$^{-1}$ is determined.

The fact that methanol does not co-desorb with either molecule sets an upper limit on low temperature methanol: CO or \ce{CH4} co-desorption. Also, the fact that most of these experiments make use of a heating rate of 10 K min$^{-1}$ should be considered as an extra constraint. At lower heating ramps the volatile molecules have more time to diffuse out of the ice, whereas (monomeric) methanol is able to form clusters and remain on the surface. If there is any interaction between CO-\ce{CH3OH} when CO desorbs to the gas-phase, it must be weak, resulting in low quantities of desorbing methanol that are below our detection limit.

The weak interaction between CO and methanol is also seen in small quantities (a few ML) of CO trapped in methanol ice after CO desorption. Depending on the amount of methanol that was used in a specific experiment, CO signals are found in IR spectra of the ice to temperatures as high as 90~K. However, the outgassing of CO does not release methanol with it. Perhaps the hydrogen bonding network that methanol forms with itself is too strong to be broken by the desorption of low volatility molecules. For the case of monomeric methanol it is possible that time scales, even at a heating rate of 10 K min$^{-1}$, are sufficiently long for methanol to form hydrogen bonds and not to co-desorb.

\section{Modelling}
\label{sec.models}

\subsection{Astrochemical model description}

The upper limits derived from the CO co-desorption experiments for \ce{CH3OH}:CO ice mixtures are implemented in an astrochemical model. Our aims are to explore whether this process, although found to be inefficient in the laboratory, is able to release a small amount of methanol into the gas-phase under astrophysical conditions. 

At the low temperatures found in dark clouds ($\sim$10~K), gas-phase methanol is routinely detected with an abundance $\sim10^{-10} - 10^{-8}$ relative to molecular hydrogen \citep[e.g.][]{bacmannfaure2016}. Cold methanol has also been detected towards so-called photon-dominated regions (PDRs) with an appreciable abundance \citep[$\sim10^{-10} - 10^{-9}$ with respect to \ce{H2},][]{guzman2013,cuadrado2017}.  
The presence of gas-phase methanol in a PDR appeared to support the theory that intact methanol could be released via the process of photodesorption \citep{oberg2009a,guzman2014}. As discussed here, this is now contradicted by recent laboratory work \citep{bertin2016,cruzdiaz2016}. 

The recent detection of cold gas-phase methanol in a protoplanetary disk provides new impetus to consider alternative desorption mechnisms \citep[][]{walsh2016}. Modelling of the CH$_{3}$OH gas-phase line profile suggests that the methanol resides in a ring with the emission peaking at $\approx 30$~au. This radius is within 10~au of the position of the CO snow line in this source \citep[][]{qi2013,schwarz2016,oberg2017,zhang2017,vanthoff2017}. The derived abundance is low, ranging from $\sim 10^{-12} - 10^{-11}$ relative to \ce{H2}, depending on the assumed vertical location of the molecule. The detected methanol transitions have upper energy levels ranging from 22 to 38~K consistent with the gas-phase methanol arising in a relatively cold region of the disk. At this radius ($\approx$~30~au) these temperatures are only reached in TW Hya below $z/r \approx 0.1$, that is, the disk midplane. Here $z$ is the disk height and $r$ the disk radius. The spatial coincidence of \ce{H2CO} and \ce{CH3OH} emission in TW Hya \citep{walsh2016,oberg2017} supports the hypothesis of a CO-ice-mediated chemistry in the vicinity of the CO snow line (and snow surface) in protoplanetary disks. Hence, protoplanetary disks offer a good test case for the proposed co-desorption route to gas-phase methanol.  

Here, we have explored the relative efficiencies of the various non-thermal desorption mechanisms proposed for non-volatile molecules like methanol at low temperatures: photodesorption, reactive desorption, and the process discussed here; thermally induced co-desorption. The chemical model used includes gas-phase chemistry and gas-grain interactions (i.e. adsorption and desorption), as well as grain-surface chemistry. The base network has been used in numerous studies of protoplanetary disk evolution and formation \citep[][and references therein for full details]{walsh2014a,walsh2015,drozdovskaya2016}. The network used here has been updated to account for new grain-surface formation pathways to glycolaldehyde and ethylene glycol \citep{fedoseev2015b,chuang2016}. The photodesorption pathways for methanol fragmentation and desorption upon UV irradiation are included \citep{bertin2016}. In addition, binding (desorption) energies of all ice species in the network have been reviewed and updated according to the literature compilation presented in \citet{cuppen2017}. However, in order to better explore the effects of the different non-thermal desorption mechanisms and simplify the grain-surface chemistry, we did not include ice photodissociation throughout the bulk ice mantle except for the case of fragmentation upon photodesorption which is restricted to the top two monolayers \citep[see e.g.][for details]{walsh2014a}. We allowed quantum tunnelling for surface reactions involving atomic and molecular hydrogen (assuming a barrier width of 1\AA), and allowed efficient diffusion of surface species at low temperature ($E_\mathrm{diff}/E_\mathrm{des}$=0.3). However, we did not consider reaction-diffusion competition; hence, the surface reaction rates are dictated by the rates of thermal and quantum hopping and the barrier height for reaction. We also only allowed surface chemistry to happen in the top two monolayers of the ice mantle.

We modelled the chemical evolution in time across vertical slices of a protoplanetary disk using a physical model representative of the disk around TW Hya \citep[from][]{kama2016}. The initial abundances of primary C-, O-, and N-containing volatiles are the same as the `molecular' set used by \citet{eistrup2016} which are representative of ice abundances in the ISM \citep{oberg2011,boogert2015}. However, we used a depleted value for sulphur (S/H $= 8.0 \times 10^{-8}$), initially in the form of \ce{H2S} ice, because \citet[][]{eistrup2016,eistrup2017} find that a high abundance of volatile sulphur (S/H $\sim 10^{-5}$) can influence the oxygen chemistry in the ice mantle. This value is more in line with observations of S-bearing species in molecular clouds and protoplanetary disks that suggest that $>99$\% of sulphur is locked up in refractory form and thus depleted from the gas and the ice phases \citep[see e.g.][for recent work on this.]{neufeld2015,guilloteau2016}. The high value used in \citet[][]{eistrup2016,eistrup2017} was to ensure that the chemical models in that work used the same initial chemical conditions as in the planet population synthesis models with which that work was comparing: the depleted value is thus more realistic. We have also included additional elements \citep[with initial abundances from][]{mcelroy2013} which are important for the ionisation balance in the disk atmosphere: Si, Fe, Na, and Mg. The chemistry is then evolved at each grid point for $\approx 10^{7}$~yr to allow extraction of abundances as a function of time up to the estimated age of TW Hya ($\sim 10$~Myr).  
 
\begin{figure*}
\centering
\includegraphics[width=\textwidth]{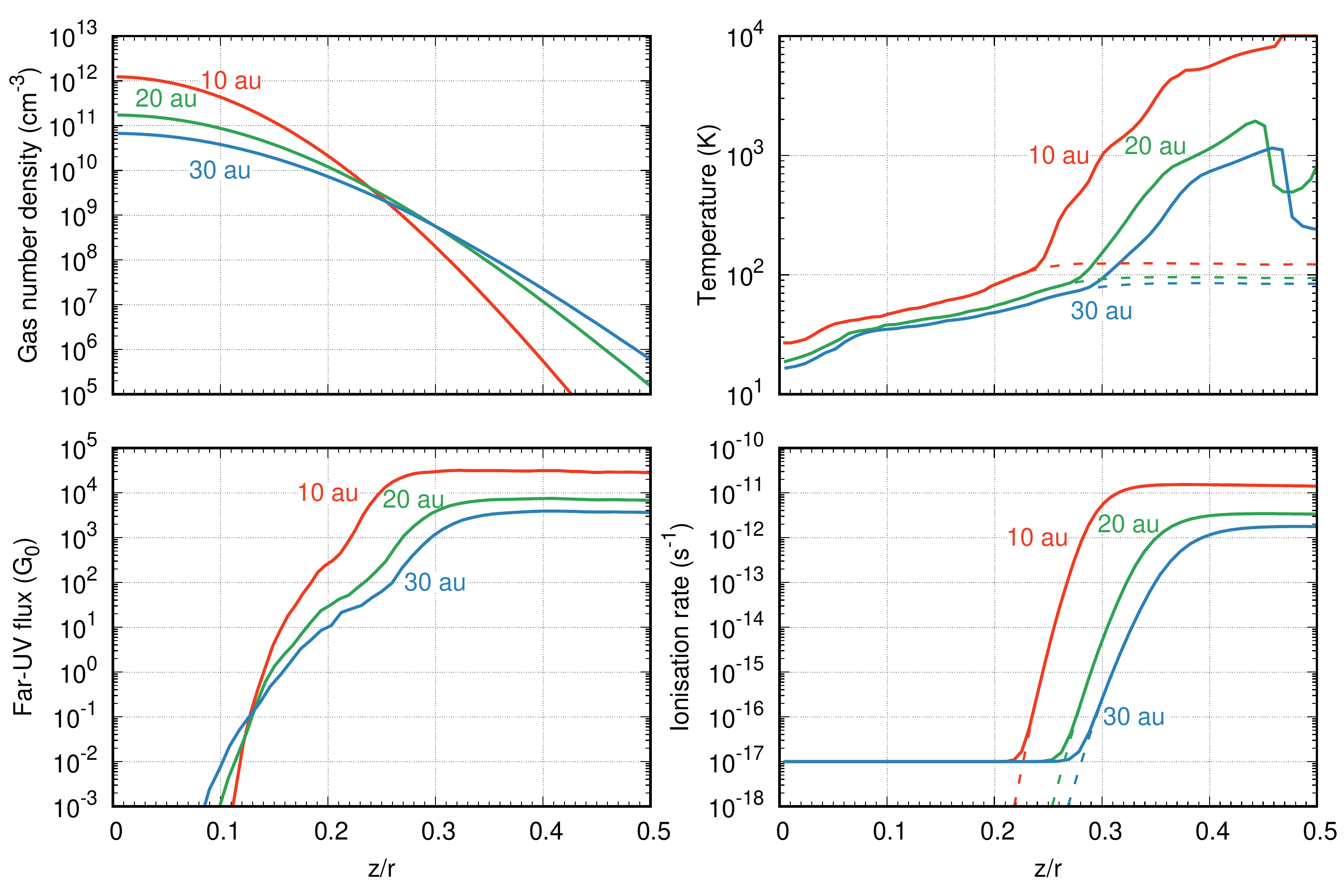}
\caption{Physical structure of the TW Hya disk slices at 10~au (red lines), 20~au (green lines) and 
30~au (blue lines) as a function of disk height, $z$, scaled by the 
radius, $r$.  
From the top-left panel and moving clockwise: gas number density (cm$^{-3}$), 
gas (solid lines) and dust (dashed lines) temperaure (K), ionisation rate (s$^{-1}$) 
due to X-rays and cosmic rays (solid lines) and x-rays only (dashed lines), and 
far-UV flux (in units of G$_0$, where G$_0 = 1.3 \times 10^{-3}$~erg~cm$^{-2}$~s$^{-1}$ is the average 
interstellar radiation field).}
\label{fig5}
\end{figure*}

\subsection{Astrochemical model results}

Figure~\ref{fig5} shows the physical structure of three vertical slices in the disk at 10, 20, and 30~au, chosen to probe the region in the vicinity of the CO snowline in TW Hya. The gas density ranges from $\sim 10^{11} - 10^{12}$~cm$^{-3}$ in the disk midplane ($z/r=0$) to $\sim 10^{6} - 10^{7}$~cm$^{-3}$ in the disk atmosphere. The gas temperature also spans several orders of magnitude across the vertical extent of the disk, from $\approx 20-30$~K in the midplane to $\gtrsim 1000$~K in the upper surface layers of the disk. Because of the diffuse nature of the gas which decreases the efficiency of gas cooling via gas-grain collisions, the gas and dust temperatures decouple such that the dust temperature remains $\lesssim 100$~K at 10 and 20~au, but the gas temperature is very high \citep[see e.g.][]{bruderer2012,kama2016}. As the primary source of heating, the far-UV flux closely follows the gas temperature: the disk midplane is significantly shielded from the stellar radiation. At all three radii considered here, the UV flux is less than the average interstellar radiation field (far-UV flux equals $1.0\ G_0$) below $z/r \approx 0.15$. In the disk surface layers, due to the proximity to the central star, the far-UV flux reaches values $\gtrsim$ a few times $10^{3} \times G_0$. 

\begin{figure*}
\centering
\includegraphics[width=\textwidth]{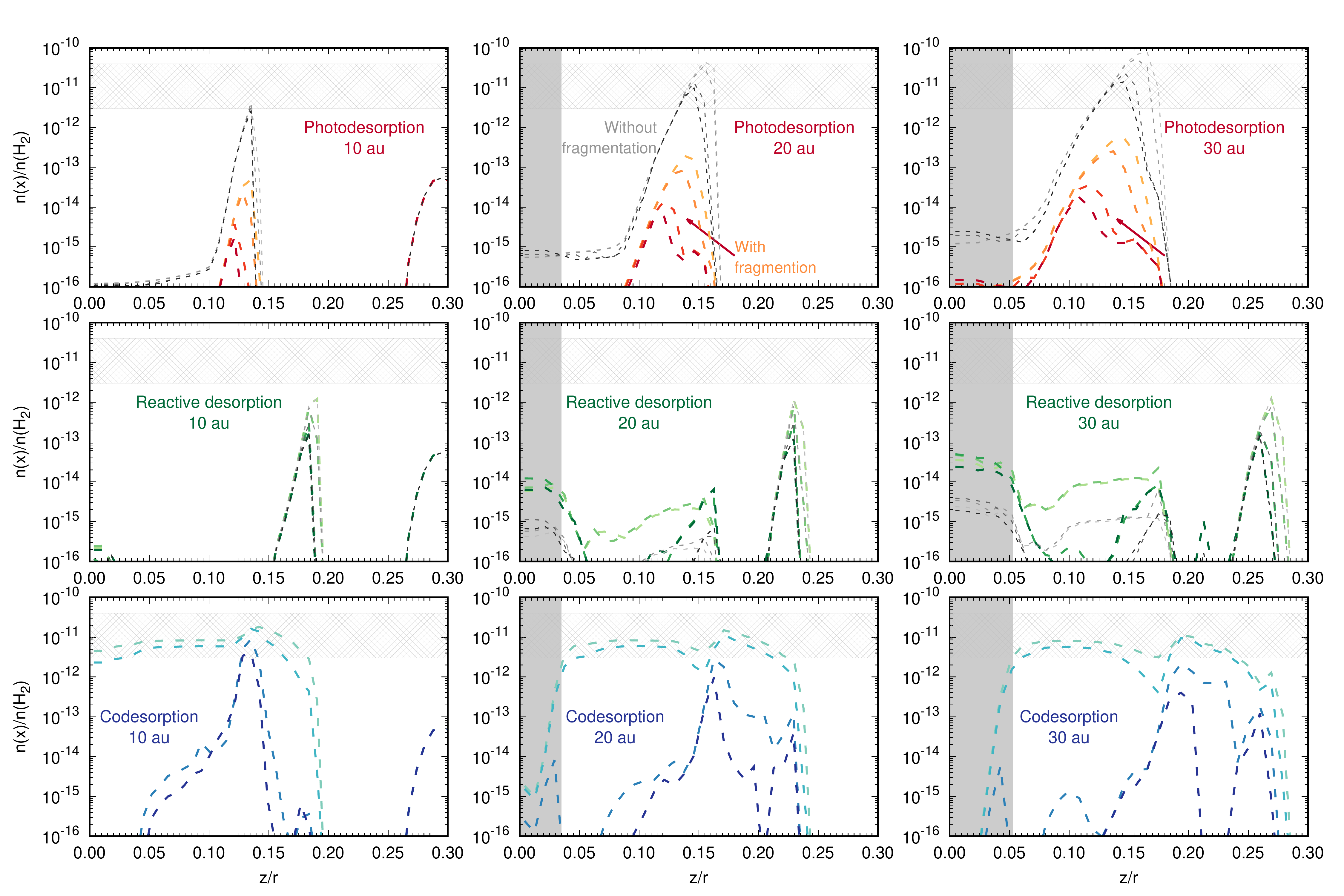}
\caption{Fractional abundance of methanol gas as 
a function of $z/r$ for radii of 10 (left), 20 (middle), and 30 (right) au.  
Results are shown for three different non-thermal desorption mechanism: 
photodesorption only (top), reactive desorption only (middle), and 
codesorption with CO only (bottom).  
The colour gradient from light to dark represents four different 
time steps: 0.5, 1.0, 5.0, and 10 Myr. 
The grey lines in the top and middle plots represent the `fiducial' model 
as described in the text
(i.e. photodesorption of intact methanol with a yield of $\sim 10^{-3}$ and a reactive desorption 
probability of 1\%). 
The vertical grey lines in the plots at 20 and 30 au mark the CO snow surface 
(see Figure~\ref{fig7}). The horizontal bar indicates the methanol abundance as observed towards TW Hya by \citet{walsh2016}.}
\label{fig6_gas}
\end{figure*}

\begin{figure*}
\centering
\includegraphics[width=\textwidth]{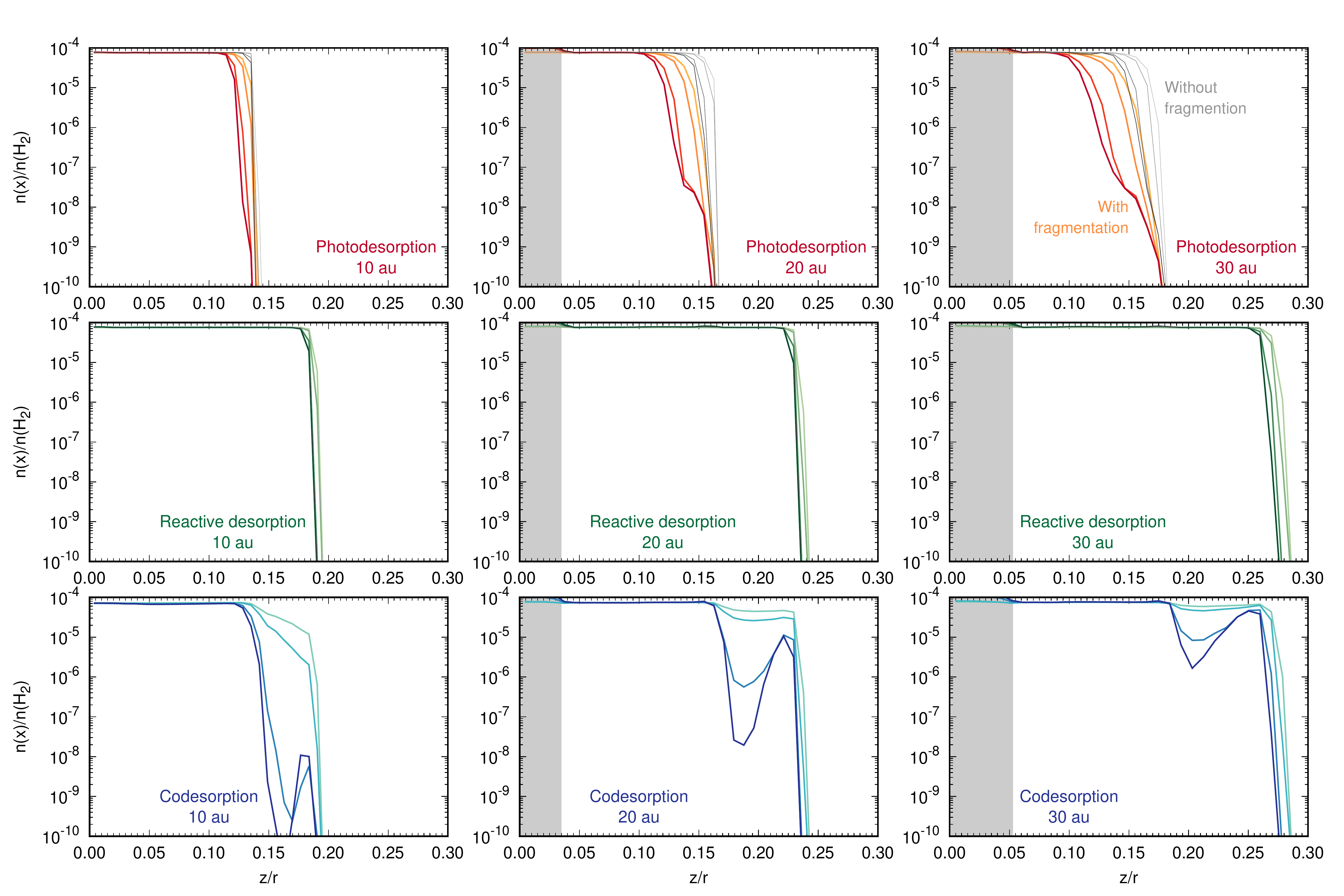}
\caption{As Fig. \ref{fig6_gas}, but for methanol ice.}
\label{fig6_ice}
\end{figure*}

Figure~\ref{fig6_gas} shows the fractional abundance of \ce{CH3OH} gas, and Fig.~\ref{fig6_ice} shows the ice as a function of $z/r$ for the explored non-thermal desorption scenarios at 10 (left), 20 (middle), and 30 (right)~au. In all plots, increasing time steps from 0.5, 1.0, 5.0, and 10.0~Myr are shown. The top panels show the results from a model with photodesorption only (PD model). Results assuming that \ce{CH3OH} does not fragment upon photodesorption \citep[i.e. assuming the rate determined by][]{oberg2009a} are plotted, as are the results for a model with fragmentation upon photodesorption. The middle row shows results from a model with reactive desorption only (RD model) at an efficiency of 1\% and at an efficiency of 10\%. This range is chosen to explore that constrained by recent analyses of reactive desorption efficiencies for a range of reactions \citep[e.g.][]{minissale2016}. The bottom panels show the results when including only the proposed co-desorption mechanism (CD model) assuming the upper limit co-desorption rate of $10^{-6}$ \ce{CH3OH} molecules per CO molecule indicated by the experiments. That is, only one \ce{CH3OH} molecule co-desorbs intact for every $10^{6}$ CO molecules which are thermally desorbed from the ice mantle.

We make several general observations. First, in the models with photodesorption only (PD model), the methanol snow surface lies deeper in the disk than for the model with reactive desorption only (RD model; $z/r < 0.2$ versus $z/r \gtrsim 0.2$). The position of the snow surface for the co-desorption only results (CD model) is intermediate between the PD and RD results. This illustrates the importance of photodesorption is setting the location of the snow surface in the disk atmosphere for non-volatile ice species such as \ce{CH3OH} and \ce{H2O} even at radii as close in as $10-30$~au. For the PD model, the position of the snow surface also becomes deeper with time. Including the fragmentation of methanol upon photodesorption has two effects: (i) the peak abundance of gas-phase methanol has decreased by two to three orders of magnitude in line with the magnitude of the decrease in the \ce{CH3OH} photodesorption rate, and (ii) the shift in the location of methanol snow surface is larger (by a factor of two or more).  This latter effect is due to the necessity for methanol ice to reform from its fragments upon photodesorption. In the original treatment, photodesorption mainly competes with re-adsorption of intact methanol.  

We now turn to the gas-phase methanol abundance and distribution, the peak in the absolute abundance coincides with the position of the methanol snow surface for both the PD and RD models. Furthermore, in both of these models the peak abundance lies always below $10^{-12}$ with respect to molecular hydrogen. For the RD model, an optimistic reactive desorption probability of 10\% is necessary to reach these values. The results for 1\% lie an order of magnitude below this. In the model with the original (and incorrect) treatment for methanol photodesorption, the abundance does reach that derived from the ALMA observations. Hence, these results show that, with a realistic treatment of methanol ice photodesorption and an optimistic reactive desorption probability, neither of these processes alone can explain the observed abundance in TW Hya.  

\begin{figure*}
\centering
\includegraphics[width=\textwidth]{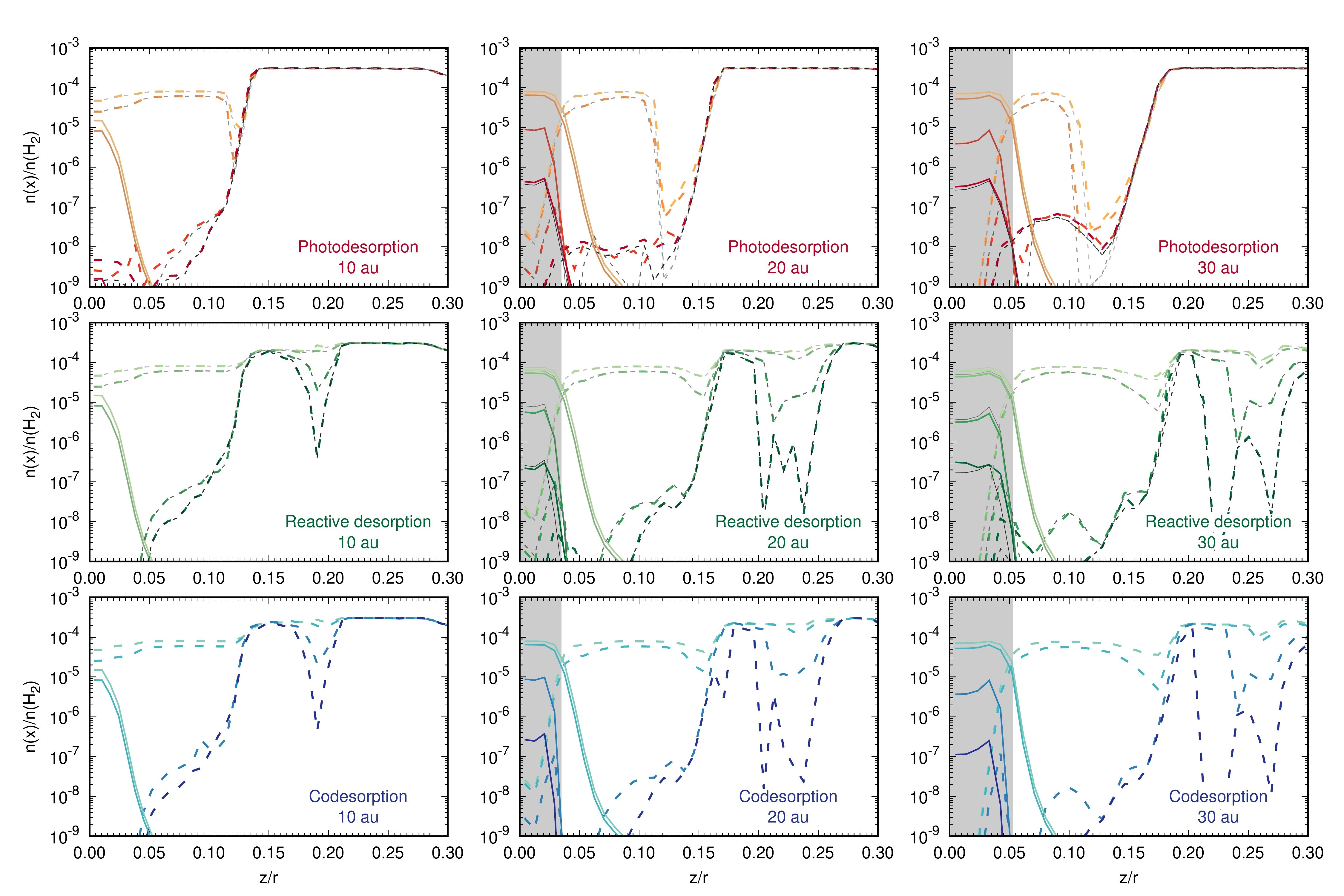}
\caption{As Fig. \ref{fig6_gas}, but for CO gas (dashed lines) and ice (solid lines).}
\label{fig7}
\end{figure*}

On the other hand, the co-desorption results using the laboratory upper limit reveal an interesting distribution and abundance for gas-phase methanol. As expected from the hypothesis of a CO-mediated ice chemistry, the gas-phase methanol peaks between the CO and \ce{CH3OH} snow surfaces covering a larger vertical extent than the other models. At early times (up to 1~Myr) the peak fractional abundance is $\approx 10^{-11}$ with respect to \ce{H2} and remains fairly constant between the two snow surfaces. However, the distribution does change with time, leading to a distribution which again peaks at around the location of the \ce{CH3OH} snow surface. The reason for this change in gas-phase methanol distribution over time is related to the concurrent chemical processing and loss of CO ice (and gas).  Figure~\ref{fig7} shows the results for CO gas and ice for the same models as shown in Figs.~\ref{fig6_gas} and \ref{fig6_ice} for methanol. Despite this model not including the radiation processing of the bulk ice mantle except for fragmentation upon photodesorption, (i.e. only gas-phase processing is included), beyond 1~Myr, CO ice and gas are significantly depleted within and around the location of the methanol snow surface. This chemical processing is one explanation for the low disk masses derived from CO observations as the canonical gas-phase ratio CO/\ce{H2} $\sim 10^{-4}$ no longer holds \citep[e.g.][]{walsh2015,reboussin2015,yu2016,eistrup2017}. Hence, co-desorption appears to only work up to 1~Myr which is the timescale within which CO is not significantly depleted. This timescale conflicts with (i) the estimated age of TW Hya (up to 10~Myr), and (ii) observational evidence for CO depletion in TW Hya \citep[e.g.][]{favre2015,schwarz2016,zhang2017}

These model runs were performed assuming the canonical cosmic-ray ionisation rate of $10^{-17}$~s$^{-1}$: this is the main source of processing in the disk midplane and is what drives the chemical conversion of CO, whether in the gas, or in the ice. Figure~\ref{fig5} shows how the X-ray ionisation level drops below the canoncial cosmic-ray ionisation rate below $z/r \lesssim 0.2 - 0.3$, depending on radius. There has been recent arguments that T Tauri stars have sufficiently strong stellar winds to deflect galactic cosmic rays \citep[evidenced by modelling of the cation abundance and emission from TW Hya;][]{cleeves2015}. Hence, if X-rays and short-lived radionuclides are the only sources of ionisation in the TW Hya disk midplane, this may (i) increase the longevity of the co-desorption effect, and (ii) delay the onset of CO depletion to better match the magnitude seen in TW Hya (around a factor of 100).  

\section{Conclusions}
\label{conclusion}

This work presents an experimental and modelling investigation of the low temperature methanol co-desorption mechanism. Laboratory experiments put upper limits on the thermal methanol co-desorption with CO. Modelling based on upper limits of these experiments shows however that low temperature methanol co-desorption can still be a significant mechanism, even at this low limit. The main conclusions of this work are summarised as follows:

\begin{enumerate}
\item Methanol is not seen co-desorbing with CO at a laboratory temperature of 30~K giving a general upper limit of $R_\mathrm{methanol} <$ a few $\times 10^{-6}$, with the lowest limit found at $R_\mathrm{methanol}~<~7.3~\times 10^{-7}$~\ce{CH3OH}~CO$^{-1}$ .
\item Co-desorption of methanol with methane is not seen down to a level of $R_\mathrm{methanol}~<~3.2~\times 10^{-5}$~\ce{CH3OH}~\ce{CH4}$^{-1}$ 
\item Results presented in this paper suggest that the interaction between desorbing CO/\ce{CH4} and solid methanol is weak and can not (easily) overcome the hydrogen bonded network of methanol.
\item Astrochemical models employing a co-desorption upper limit of 10$^{-6}$ \ce{CH3OH} CO$^{-1}$ are able to reproduce the observed abundance and distribution of gas-phase methanol between 10-30 au in the TW Hya protoplanetary disk. The models employing a more realistic treatment of methanol photodesorption only or an optimistic treatment of reactive desorption only, do not reproduce the observed abundance. We note, though, that the actual thermal co-desorption may be less efficient, as the experiments discussed here only provide upper limits.  
\item The gas-phase methanol peaks between the CO and \ce{CH3OH} snow surfaces;  yet, a moderate abundance ($\sim$10$^{-11}$) is retained only up to $\sim$1 Myr, beyond which chemical processing of CO impedes the co-desorption effect.  
\item The chemical processing of CO in the disk midplane is driven primarily by cosmic rays: it remains to be tested whether processing by X-rays and/or short-lived radionuclides can help the co-desorption effect persist to the estimated age of TW Hya ($\sim$10 Myr).
\item Although the models suggest that thermal co-desorption could contribute to the production of gas-phase methanol at and around the CO snowline in protoplanetary disks, the lack of a confirmed signal in the laboratory experiments means that the impact of the actual thermal co-desorption may be less, as the experiment only provides an upper limit. As a consequence, other non-thermal desorption mechanisms cannot be ruled out at this time.
\end{enumerate}

\begin{acknowledgements}
The authors of this paper would like to thank Ewine van Dishoeck for useful input and discussions on this work. Astrochemistry in Leiden is supported by the European Union A-ERC grant 291141 CHEMPLAN, by the Netherlands Research School for Astronomy (NOVA) and by a Royal Netherlands Academy of Arts and Sciences (KNAW) professor prize. CryoPAD2 was realised with NOVA and NWO (Netherlands Organisation for Scientific Research) grants. C. Walsh acknowledges financial support from the Netherlands Organisation for Scientific Research (NWO, programme 639.041.335) and start-up funds from the University of Leeds.
\end{acknowledgements}


\bibliographystyle{aa}
\bibliography{lib}

\end{document}